\documentclass[twocolumn,floatfix]{revtex4}
\usepackage{xcolor}
\usepackage{epsfig}
\usepackage{graphicx}
\usepackage{amsmath}

\newcommand {\be}  {
\begin{equation}
}
\newcommand {\ee}  {
\end{equation}
}
\newcommand {\bea} {
\begin{eqnarray}
}
\newcommand {\eea} {
\end{eqnarray}
}

\newcommand{\RN}[1]{%
  \textup{\uppercase\expandafter{\romannumeral#1}}%
}
\newcommand{\ignore}[1]{}

\begin{document}

\title{Pattern formation during melting of lamellar eutectics}
\author{Rahul Nellissery Rajan$^1$, Rajesh Kumari Rajendran$^2$, Guillaume Boussinot$^1$, Kamal Sbargoud$^2$, Sabine Bottin-Rousseau$^2$, Silv\`ere Akamatsu$^2$}
\affiliation{$^1$Access e.V., Intzestr. 5, 52072 Aachen, Germany}
\affiliation{$^2$Sorbonne Universit\'e, CNRS, Institut des Nanosciences de Paris, 75005 Paris, France}

\begin{abstract}

We present a study of the melting dynamics of a two-phase eutectic solid. In situ, thin-sample experiments using a transparent eutectic alloy and two-dimensional phase field simulations calibrated for the very same alloy are combined to assess pattern formation during directional melting in a temperature gradient. Depending on the melting velocity $V_m$ and the spacing $\lambda$ of the pre-solidified  lamellar microstructure, an unexpectedly rich diversity of melting patterns is observed, with good agreement between experiments and  simulations. We unravel the different physical mechanisms leading to this diversity, and establish the scaling behaviors of (i) the penetration of the liquid along the solid-solid interface at large $V_m$, (ii) the thickening of the primary-phase fingers at low $V_m$, and (iii) a period-doubling instability  for small $\lambda$ values. Our study provides a fundamental basis for further investigations of eutectic melting, including additive manufacturing during which melting/solidification cycles take place.
\end{abstract}

\maketitle

Melting plainly refers to the reverse process of solidification. Nevertheless, the way a material melts upon heating is very different from the way crystals grow from a liquid upon cooling. Indeed, while solidification microstructures arise from self-organizing patterns at the growth front, the very same microstructures serve as an initial condition for melting. This poses deep questions of key relevance to natural phenomena \cite{Schmidt2023,Xu2002} and elaboration processes \cite{RMP_melting}. Recently, this largely unexplored research topic started to attract more attention  in reference to additive manufacturing, during which a material is repeatedly submitted to partial melting and solidification cycles \cite{ghosh, korner}. In this respect, multiphase eutectics are specially attractive. 
Eutectic solidification delivers composite microstructures with remarkable  properties \cite{Tiwary2022,Kang2024}, including via additive manufacturing \cite{Plotkowski2017,materials_design, moi_protrusion}. 
It is also a paradigmatic nonequilibrium pattern formation phenomenon  \cite{Langer80,hohenberg,faivre_tilt,kassner_misbah,karma_sarkissian,ginibre,AkamatsuPlapp,folch,spiral, moi_eutectic}. 
In contrast, eutectic melting has still received small attention on a fundamental level  \cite{nordlung_trivedi,brener_temkin,BrenerBoussinot2009,granasy}. 
We propose an investigation of the melting of eutectic microstructures by combining in situ experiments using thin samples of a model  alloy, and two-dimensional phase-field simulations calibrated for that very same alloy.  
The aim is to cast new light on physical mechanisms that govern   steady-state melting patterns and their morphological transformations.

\begin{figure}[htbp] 
\begin{center}
\includegraphics[width=7.7cm]{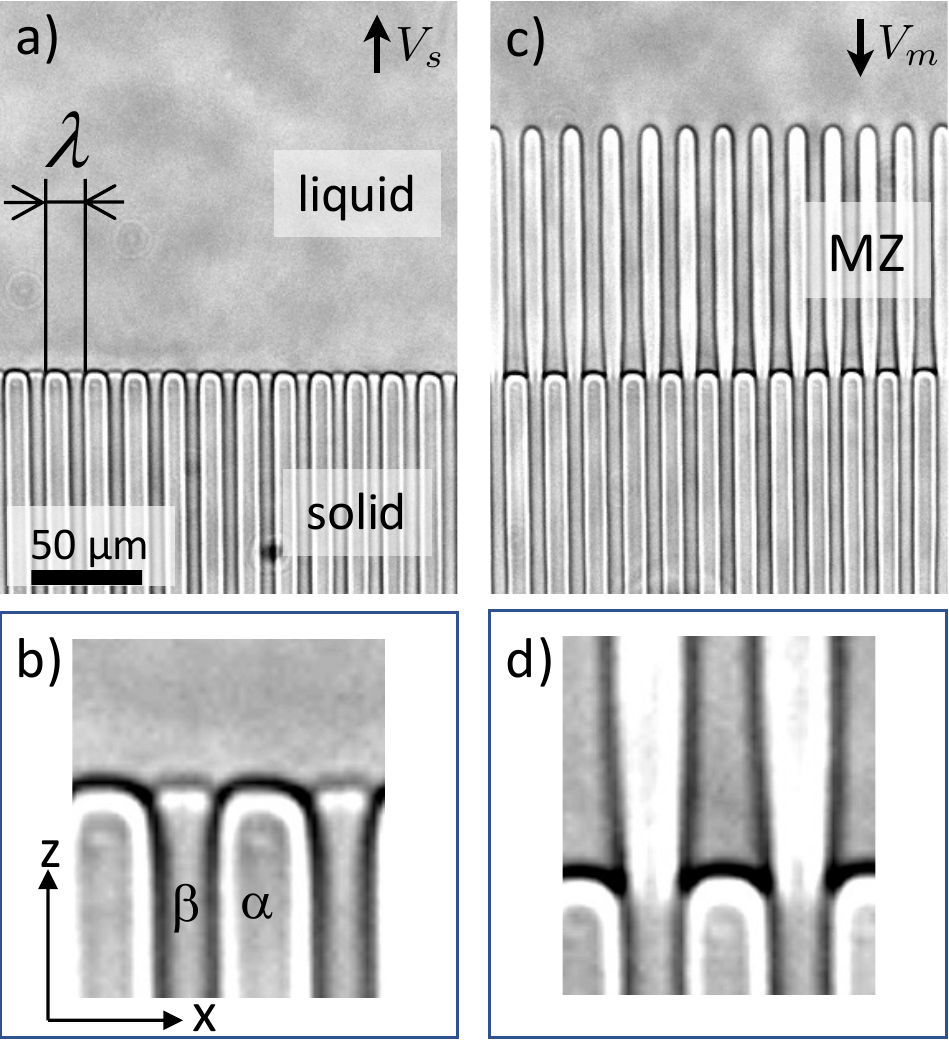}
\end{center}
\caption{In situ optical images in a thin CBr$_4$-C$_2$Cl$_6$ sample. a) and b) Directional solidification of a lamellar growth pattern ($V_s = 1~{\rm \mu ms^{-1}}$). c)  and d) Directional eutectic melting pattern ($V_m = 1~{\rm \mu ms^{-1}}$) issued from the microstructure formed in a). b) and d) are 4-times enlarged details of the region of the trijunctions for the patterns in a) and c), respectively. MZ: $\beta$-liquid coexistence region (mushy zone). $\alpha$ and $\beta$: eutectic solid phases. {\bf z}: axis of the temperature gradient ({\bf x} is parallel to the isotherms). $\lambda$: interlamellar spacing.}
\label{solmelt}      
\end{figure}

In a binary eutectic, two solid phases  $\alpha$ and $\beta$  coexist with the liquid at the eutectic temperature $T_E$. Upon cooling, the liquid solidifies into $\alpha$$\beta$ microstructures. Good control of their spatial distribution can be achieved in directional solidification (DS) of a nonfaceted alloy at an imposed velocity $V_s$ in a fixed temperature gradient $G$. Regular (e.g. lamellar) microstructures are then delivered by periodic, coupled-growth patterns that extend in steady-state over a planar isotherm close to $T_E$. 
Depending on boundary and initial conditions, the spacing $\lambda$ of a lamellar eutectic (Figs. \ref{solmelt}a and \ref{solmelt}b) falls close to a characteristic length, the Jackson-Hunt spacing $\lambda_{\text{JH}}$, which varies as $\lambda_{\text{JH}} \sim V_s^{-1/2}$ \cite{JH,kassner_misbah}. The physics of the system is determined by solute diffusion in the liquid on the scale of $\lambda$, and capillary (or curvature) effects. This basic description holds for a wide range of alloy concentrations around the eutectic point, independently of $G$, basically owing to the overall planarity of the coupled-growth front.

For directional melting (DM), it suffices to change the sign of the sample translation speed in the temperature gradient. Yet, as illustrated in Figs. \ref{solmelt}c and \ref{solmelt}d, eutectic melting patterns are, after an initial transient (see movies in Supplemental Material), commonly far from being planar, and a new theoretical approach has to be found. 

The nonfaceted transparent CBr$_4$-C$_2$Cl$_6$ alloy (Table \ref{table1}) is a model  lamellar eutectic that has been previously used for numerous solidification studies (see, e.g.,  \cite{mergy,ginibre}), and  for eutectic melting in Ref.  \cite{nordlung_trivedi}. The chosen nominal concentration $C_0$ is  slightly hyper-eutectic  ($C_0>C_E$). In the solid, the $\beta$ ($\alpha$) phase  has a volume fraction of about 0.35 (0.65). Thin  (12~$\mu$m-thick) samples are protected by flat glass walls.  In the DS setup, a thermal gradient ($G=9\pm 1~\rm{Kmm^{-1}}$) establishes by heat diffusion in the sample (axis {\bf z}) in contact with two metallic blocks regulated at temperatures $T^+$ and $T^-$ ($T^+ > T_E > T^-$; see Fig. \ref{TDS}). Directional solidification or melting is carried out by translating the sample along {\bf z} towards either the $T^-$ or the  $T^+$ block. The DS setup is placed on the XYZ stage of an optical, transmitted-light microscope equipped with a numerical camera --for more details, see Ref. \cite{AkamatsuLeeLosert}.  For numerical simulations, we used the well-known phase-field (PF) method \cite{folch}. In the PF study of Ref. \cite{granasy}, the authors considered a simplified symmetric eutectic alloy.  Here, we used the MICRESS  software \cite{micress} for quantitative simulations implementing the physical parameters of  the  CBr$_4$-C$_2$Cl$_6$ system. We assume vanishing diffusion in the solid phases and local equilibrium at the solid-liquid interfaces.  In the simulations, we set $G=22~\rm{Kmm^{-1}}$ for minimizing the size of the simulation box along {\bf z}, thus sparing computation time. We took the contact angles of the $\alpha$- and $\beta$-liquid interfaces at a triple junction (trijunction) both equal to 60$^\circ$  without loosing accuracy with respect to experimental uncertainties.

\begin{table}[htbp] 
   \begin{center} 
\begin{tabular}{|c |c  |c |c |c | c | c |c |} \hline
$T_E$ &  $C_E$  &$C_\alpha$ & $C_\beta$ & $m_{\beta}$ & $D$ & $d_0$ & $C_0$ \\
  $[K]$ &  [mol\%] &  [mol\%]  &  [mol\%] &  [K/mol]  & [${\rm \mu m^2s^{-1}}$] & [${\rm \mu m}$] &   [${\rm mol\%}$] 
\\\hline
357.4  & 11.6 & 8.8 & 18.5 & 165 & 500 & 0.011 & 12.2 
\\\hline
\end{tabular}
  \caption{Parameters of the CBr$_4$-C$_2$Cl$_6$ alloy \cite{mergy}. $T_E$: eutectic temperature; $C_E$ eutectic concentration; $C_\alpha$: concentration of the eutectic solid $\alpha$ at $T_E$;$C_\beta$: concentration of the eutectic solid $\beta$ at $T_E$; $m_\beta$: slope of the $\beta$ liquidus; $D$: solute diffusion coefficient in the liquid; $d_0$:  capillary length; $C_0$: nominal  concentration of the alloy. \label{table1}}
   \end{center} 
\end{table}

\begin{figure}[htbp] 
\includegraphics[width=7.8cm]{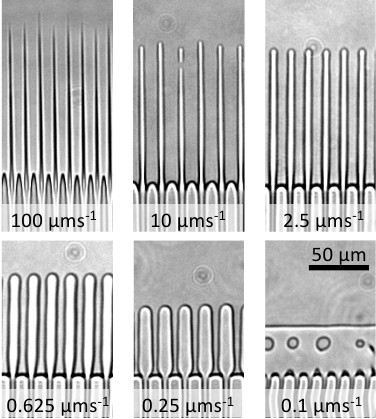}
  \caption{In situ experimental snapshots (details) of directional melting patterns for various $V_m$. Same alloy as in Fig. \ref{solmelt}. The thermal gradient is oriented vertically upwards. Here, the eutectic $\alpha$-$\beta$ structure is generated by a solidification stage at a velocity $V_s=2~{\rm \mu ms^{-1}}$.}
  \label{digging}
\end{figure}

\begin{figure}[htbp] 
  \includegraphics[width=7.9cm]{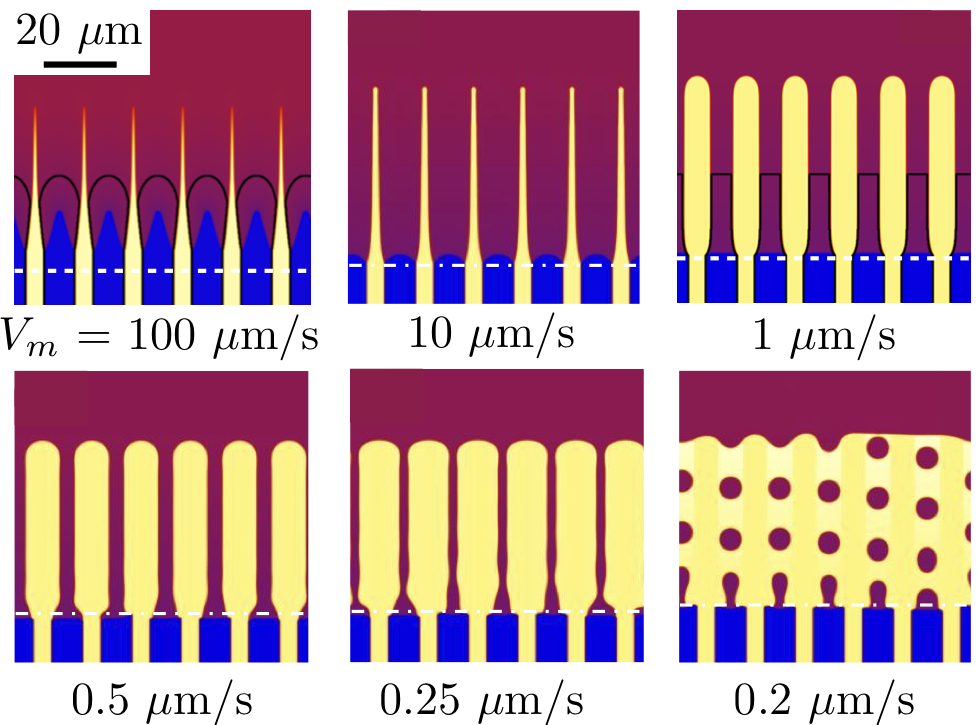}
  \caption{Two-dimensional phase-field simulations of directional melting patterns in the  CBr$_4$-C$_2$Cl$_6$ system  at varying $V_m$ ($V_s=1~{\rm \mu ms^{-1}}$). The liquid is shown in red, the $\beta$ phase in yellow, and the $\alpha$ phase in blue. The white horizontal dashed-dotted line locates the eutectic temperature, while for $V_m=100$ $\mu$m/s and 1 $\mu$m/s, an iso-concentration line is shown in black.}
  \label{digging_sim}
\end{figure}

In both experiments and simulations, we first solidify a lamellar structure at $V_s$ ($1-10~{\rm \mu ms^{-1}}$). The resulting spacing is close to $\lambda_{\text{JH}}=\sqrt{d_0 D/(V_s\mathcal{P})}$  \cite{JH}, where $d_0$ is a weighted solid-liquid capillary length, $D$ is the solute diffusion coefficient in the liquid, and $\mathcal{P} \approx 0.0293$ is an integration constant \cite{mergy}. Directional melting of  the solid microstructure is performed at a velocity $V_m$ ($0.1-100~{\rm \mu ms^{-1}}$) close to, or much different from $V_s$. 
Figure \ref{digging} shows experimental DM patterns for various $V_m$ values at a given $\lambda$.  
For comparison, PF simulations for a similar sequence are displayed in Fig. \ref{digging_sim}a.
Simulations and experiments are in a good agreement.
Overall, our numerical results  are also in  qualitative accordance with those  of Ref. \cite{granasy}.
 
The sample is fully solid below $T_\text{TJ}$, the temperature at the trijunction (TJ) (see Figs. \ref{solmelt}c and \ref{solmelt}d).
 Above $T_\text{TJ}$, $\beta$ protrudes, yielding a mushy zone (MZ) consisting of an array of apparently almost straight  $\beta$ fingers extending, in continuity with the $\beta$ lamellae in the solid, to the tip temperature $T_\text{tip}>T_\text{TJ}$.  
Then, above $T_\text{tip}$, the system is fully liquid.
Assuming a temperature-dependent Gibbs lever-rule condition, the hyper-eutectic composition $C_0$ provides $\beta$ with the role of so-called primary phase (this role is acquired by $\alpha$ when $C_0$ lies below $C_E$ in the hypo-eutectic case), and one may at first glance understand in this way the protrusion of $\beta$.
However, as shown later, a primary solid phase melting necessarily at a higher temperature than the other solid phase should not be considered as a general rule.

Pattern formation during DM of eutectics can be described in terms of a competition between several length scales involving the four control parameters $C_0$, $V_m$,  $G$, and $\lambda$, but also the capillary length $d_0$. 
While only two parameters, namely, $V_m$ and $\lambda$ vary in this study, several morphological transitions were apprehended, for example when, at low enough $V_m/V_s$, the actual periodicity of the eutectic melting pattern becomes a multiple of $\lambda$.  
Obviously, the multiplicity of characteristic length scales in competition pushes far beyond the scope of this Letter a complete unravelling of the pattern formation process during DM of eutectics. 
For example, since $C_0$ is kept constant here, the transition,  when $C_0$ becomes close enough to $C_E$, from what is called "non-coupled" to "coupled" melting in Ref. \cite{nordlung_trivedi, granasy} is  not described here.
Moreover, three-dimensional effects, imperfections of the experimental set-up (for example a degradation of the sample at long melting times), or uncertainties on physical parameters, may hinder a full correspondance between the conditions of the competition between characteristic length scales in experiments and in simulations.
As a consequence, qualitative differences between observed and simulated patterns may appear. 
For example, in contrast to the PF simulations for which we have $T_\text{tip} \approx T_0$ where $T_0=T_E+m_\beta(C_0-C_E)$ is the liquidus temperature and $m_\beta$ the liquidus slope, $T_\text{tip}$ exceeds significantly $T_0$ at large $V_m$ in experiments.


One can distinguish three different types of regions along {\bf z} for the diffusion field in the liquid: 
{\bf(i)} the neighborhood of the TJs where the three phases ($\alpha$, $\beta$, liquid) are present, with diffusion fluxes naturally generated owing to the difference between the concentrations at the $\alpha$-liquid ($C_{L\alpha}$) and the $\beta$-liquid ($C_{L\beta}$) interfaces;
{\bf(ii)} the MZ where the primary $\beta$ crystals coexist with the liquid (the $\alpha$ phase is absent), and where lateral (i.e. perpendicular to the thermal gradient) and vertical (i.e. aligned with it) fluxes compete;
{\bf(iii)} the bulk liquid, where vertical fluxes may develop, a priori on the scale of the diffusion length $l_d=D/V_m$, when $T_\text{tip}$ deviates from the liquidus temperature $T_0$.  
Note that the lateral fluxes in the MZ, that increase with $V_m$, ensure a continuous melting of $\beta$ from $T_\text{TJ}$ to $T_\text{tip}$. On the other hand, the $V_m$-independent vertical flux in the MZ, driven by the temperature dependence of $C_{L\beta}$, ensures the melting of the $\beta$ tip at the boundary between (ii) and (iii). 

Up to a small curvature effect, the temperature $T_\text{TJ}$ is very close to $T_E$.
The $\alpha$ crystals melt close to $T_\text{TJ}$, but while the $\alpha$-liquid interface is quite flat  when $V_m$ is small, it bulges significantly into the liquid when $V_m$ is large. 
In the latter case (in practice, $V_m \geq 10~{\rm \mu ms^{-1}}$), the $\beta$ fingers take the form of thin needles with sharp tips. 
Importantly, in this regime, the diffusion flux close to the TJ is perpendicular to $V_m$ and connects the $\alpha$- and $\beta$-liquid interfaces  (see Fig. \ref{TJ}a - blue frame).  The melting of the two solids is  strongly coupled, i.e. it is driven by $\Delta C_\text{TJ}=C_{L\alpha}(T_\text{TJ}) - C_{L\beta}(T_\text{TJ})$. 
The liquid penetration  dynamics along the $\alpha$-$\beta$ interphase boundary is then similar, at least locally,  to the one studied theoretically by Brener and collaborators \cite{brener_temkin,BrenerBoussinot2009}. 
This establishes when the  length $\rho = \sqrt{d_0 D/V_m}$, which  gives a measure for the radius of curvature of the interfaces close to the TJ \cite{kassner_misbah, moi_eutectic,moi_monotectic, moi_peritectic}, is much smaller than the spacing. Since $\lambda \approx \lambda_{JH}$, the condition $\rho \ll \lambda$ is equivalent to  $V_m \gg V_s$.  
This regime is clearly reached for $V_m=100~{\rm \mu ms^{-1}}$.
Then $T_\text{TJ}$ and the shape of the DM pattern in the neighborhood of the TJ are  largely independent of $\lambda$ and $G$, and the quantity $\Delta C_\text{TJ}$, proportional to $T_\text{TJ} - T_E$, scales as  the Peclet number $\rho/l_d$. 
Both $\alpha$- and $\beta$-liquid interfaces are slightly concave. The concentration in the liquid at the TJ lies above the two liquidus lines, outside the two-phase regions  (see Fig. \ref{TJ}b), and $T_\text{TJ} > T_E$. 
This yields a solute flux in the liquid from $\beta$ to $\alpha$, of opposite sign as compared to eutectic coupled growth.
Moreover, the flux within the MZ (ii) presents a significant lateral component illustrated by the curvature of the black iso-concentration line, yielding a continuous melting of $\beta$ between $T_\text{TJ}$ and $T_\text{tip}$, in line with the direction of the flux near the TJ (i).
Let us also note here that, for an even smaller ratio $\rho/\lambda$, nothing guarentees that $\alpha$ melts at a lower temperature than $\beta$, owing to the large fraction of $\alpha$ in the solid. 
This renders inoperative a hypothetical criterion stating that the solid phase melting at the highest temperature is the primary one.

\begin{figure}[htbp] 
  \includegraphics[keepaspectratio, width=0.4\textwidth]{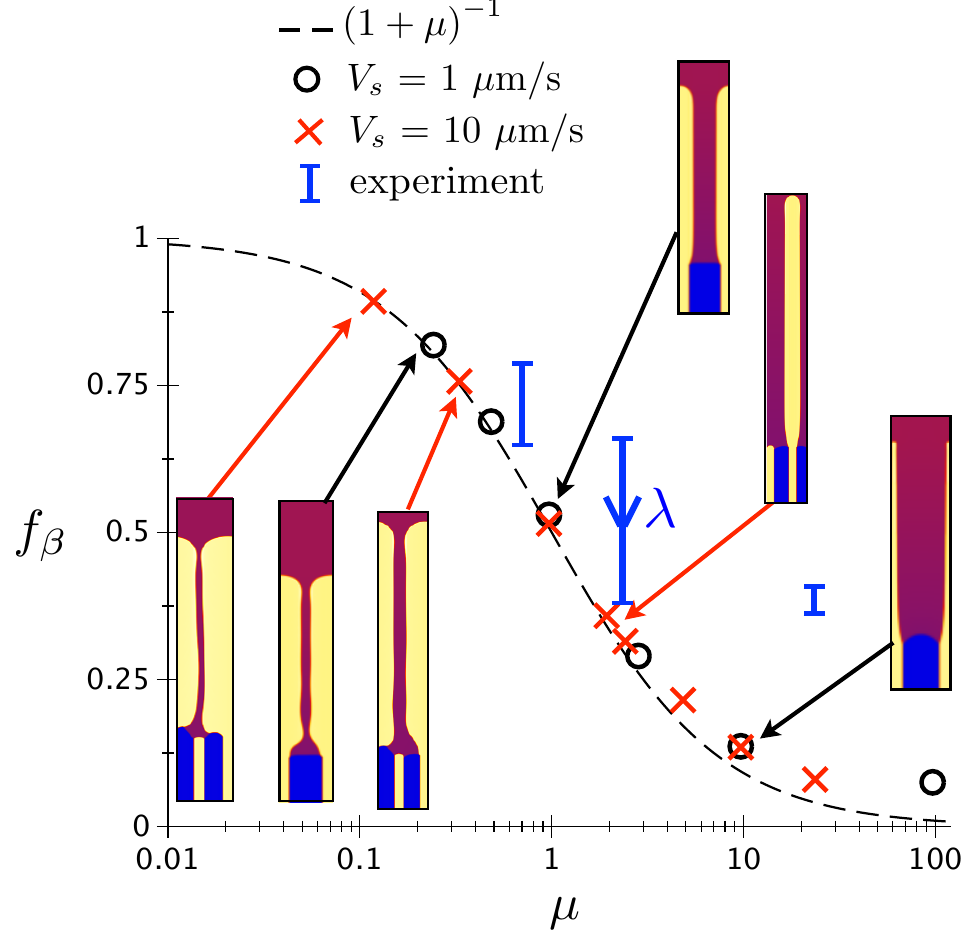}
  \caption{Fraction of $\beta$ phase $f_\beta$ in the solid-liquid (MZ) region as a function of $\mu = V_m/V_c$ (see text). Symbols: PF simulations (1-$\lambda$ and 2-$\lambda$  patterns; see below). Blue vertical lines:  experimental data  (arrow: increasing $\lambda$). Dashed line: Eq. (\ref{tau}). }
  \label{beta_frac}
\end{figure}

When $V_m$ decreases (here, below $10~{\rm \mu ms^{-1}}$), the condition $\rho \ll \lambda$ is no longer valid. The diffusion flux close to a TJ acquires a substantial component parallel to the {\bf z} axis (see Fig. \ref{TJ}a - pink frame). Moreover, the $\alpha$-liquid interface progressively becomes  significantly convex (curvature of order 1/$\lambda$), while the $\beta$-liquid interface remains slightly concave. The temperature of the TJ decreases below $T_E$, as previously noticed in Ref. \cite{granasy}, with a concentration lying inside $(\alpha+L)$ and outside $(\beta+L)$.
This still permits a solute flux from $\beta$ to $\alpha$ in the liquid, and the diffusive coupling between the two melting solids is sustained in that regime as well.
In the MZ, the $\beta$ fingers thicken, i.e., their width $l_\beta$ increases (Figs. \ref{digging} and \ref{digging_sim}) when $V_m$ decreases. This may be explained as follows.
The lateral flux progressively vanishes as illustrated by the flatness of the black iso-concentration line at $V_m=$ 1 $\mu$m/s and, when $V_m$ is small enough, the $\beta$ fingers melt only at their very tip.  
Then at steady-state, the total amount of solute transported  by diffusion in the inter-finger liquid channels of width $l_L=\lambda - l_\beta$ is balanced, for conservation of mass, by the melting  of the $\beta$ tips at velocity $V_m$.
Since the vertical flux in the MZ is approximately given by the quantity $DG/m_\beta$, this entails $l_L DG/m_\beta  \approx l_\beta V_m (C_\beta - C_0)$, 
where $C_\beta$ is taken at $T_E$ (see Table) for simplicity -- we recall that in the simulations,  the temperature of the tip of the $\beta$ finger remains approximately equal to that of the $\beta$ liquidus for $C_0$.
 Let us define  the thermal length $l_t=m_\beta (C_\beta - C_0)/G$.
 The mass conservation can then be re-written so as to find the $\beta$ fraction $f_\beta =l_\beta/\lambda$ ($2f_\beta =l_\beta/\lambda$ for 2-$\lambda$ patterns - see thereafter) in the MZ as: 
\bea\label{tau}
f_\beta = (1 + \mu)^{-1} \, ,
\eea
where $\mu=l_t/l_d =V_m/V_{c}$. We note that the characteristic velocity $V_{c}=D/l_t$ is $V_{c}= 1.06~{\rm \mu ms^{-1}}$ for the  simulations, and $\approx 0.43~{\rm \mu ms^{-1}}$ for the experiments, the difference stemming from the different values of $G$. 
As shown in Fig. \ref{beta_frac}, Eq. (\ref{tau}) predicts well the variation of $f_\beta$ in the PF simulations, evidencing the validity of the approximation  for $V_m \lesssim V_{c}$. As expected, the PF data deviate from this law for large $V_m$ when lateral fluxes are non-negligeable. 
The experimental  data reported  in Fig. \ref{beta_frac}   fall substantially above the numerical data. They also exhibit a large dispersion, at given $V_m$, which actually corresponds to a dependence of $f_\beta$ on the local spacing in slightly nonuniform microstructures. More specifically, $f_\beta$ decreases, and approaches the theoretical curve, when $\lambda$ increases. We also shall note that uncertainties relative to image processing and/or three-dimensional effects become less important for large $\lambda$ values. 

While, for $V_m$  close to  $V_c$, eutectic melting patterns are true steady-states, we observed an oscillatory motion of the TJs about their average position when $V_m \ll V_c$.  This can be interpreted as follows. For $V_m \approx V_c$,  the concentration gradient in the liquid pocket $\Delta C_\text{TJ}/\rho$  \cite{brener_temkin}  is of order $G/m_\beta$. The solute diffusion flux near the TJ  essentially matches that in the MZ, which is a condition for steady-state motion at $V_m$. 
When $V_m \ll V_c$, the diffusion gradient that is necessary for the TJ to move at $V_m$ becomes smaller than the diffusion gradient within the MZ. 
This is actually accommodated by an oscillatory dynamics, with a pulsation (alternation of growth and shrinkage) of a liquid pocket in the vicinity of the TJ, and provokes an undulation of the sides of the $\beta$ fingers (see Figs. \ref{digging} and \ref{digging_sim}).
On average, Eq. (\ref{tau}) remains nevertheless valid (Fig. \ref{beta_frac}). 
For the lowest $V_m$ values, the oscillatory dynamics is inefficient, and the $\beta$ fingers coalesce.
Melting of the primary $\beta$ phase then takes place when liquid droplets, migrating owing to the so-called thermal gradient zone melting (TGZM) phenomenon \cite{pfann, nguyen, nguyen2, boussinot_TGZM} at a velocity close to $V_c$, reach the bulk liquid. These droplets are emitted close to the TJs
- see Fig. (\ref{digging_sim}).
We note that the inaccurate treatment of small size droplets or liquid nucleation in PF simulations does not invalidate the qualitative picture.  

\begin{figure}[htbp] 
  \includegraphics[width=6.5cm]{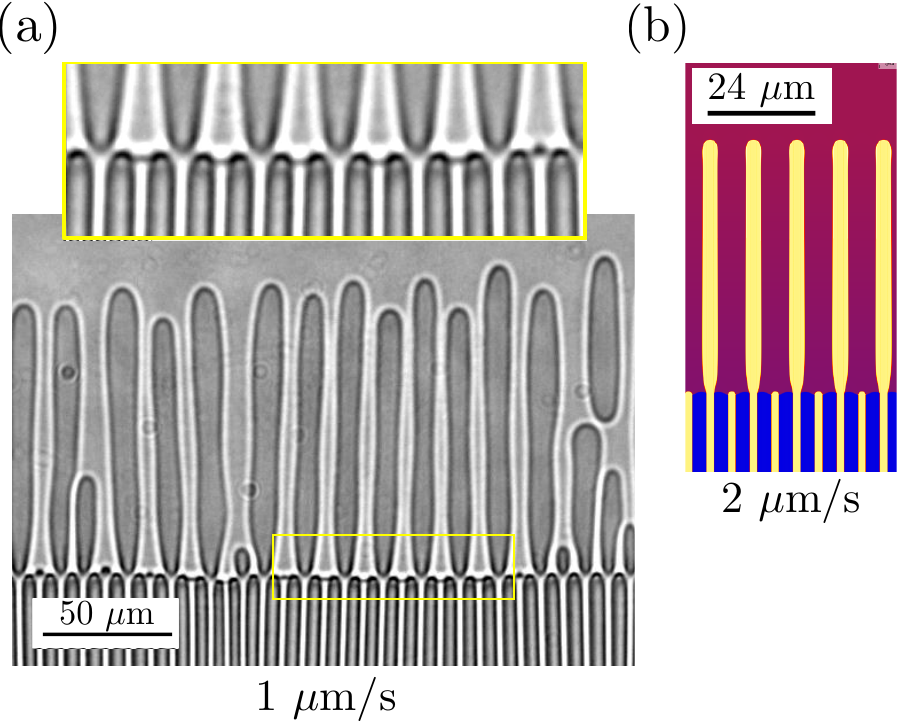}
  \caption{a) Experimental observation of a period-doubling  (2-$\lambda$)  eutectic melting pattern after  $V_m$ was decreased down to $1~{\rm \mu ms^{-1}}$ from $V_m= V_s=10~{\rm \mu ms^{-1}}$ (same sample as in Fig. \ref{solmelt}). Inset: magnified view (yellow frame). b) Phase-field simulation of a 2-$\lambda$ pattern ($V_s=10~{\rm \mu ms^{-1}}$; $V_m=2~{\rm \mu ms^{-1}}$).}
  \label{doubling}
\end{figure}

Up to this point, we have mainly considered period preserving (1-$\lambda$) melting patterns. A period doubling (2-$\lambda$) instability has also been brought to  light both experimentally and numerically  (Figs. \ref{doubling}a and \ref{doubling}c; see also Fig. \ref{beta_frac}). We consider a lamellar microstructure with a smaller spacing than above, produced by solidification at a higher velocity ($V_s=10~{\rm \mu ms^{-1}}$;  $\lambda_{JH}\approx 4.4~{\rm \mu ms^{-1}}$). For $V_m \approx V_s$,  1-$\lambda$ patterns form, in consistency with the  above results. 
In contrast, for $V_m \ll V_s$,  we observed a 2-$\lambda$ instability with one $\beta$ lamella out of two developing into a finger in the MZ, while the other $\beta$ lamella melts close to $T_E$, at the same level as the $\alpha$ lamellae. 
Preliminary PF simulations demonstrate that this transition still exists in absence of thermal gradient. This deserves more investigations in the future. 

Let us finally note that the observed morphological changes must go along with a rotation of the three interfaces at the TJ, as  in the idealized case studied in Ref. \cite{brener_temkin}. While the solid-liquid interfaces are obviously free to rotate, a bending of the $\alpha$-$\beta$ interphase boundaries, which are straight interfaces in the solid,  also entails a local curvature that is hardly observed experimentally. In the PF simulations (also see Ref. \cite{granasy}), it is enabled (and possibly exaggeratedly) by the finite diffusion across the interfaces, the diffusion coefficient varying smoothly from $D$ in the liquid to 0 in the solid  on the scale of the interface width $W$. However, we checked the independence of our results with respect to (reasonable) changes in $W$. This effect, which has been identified for a long time in PF simulations of eutectic growth (see, e.g., Ref. \cite{folch}), is still under discussion.


In conclusion, we have studied the directional melting of a regular eutectic lamellar structure, using in situ thin-sample experiments and two-dimensional phase field simulations in the model CBr$_4$-C$_2$Cl$_6$ alloy. This combined methodology allowed us to uncover the basic physical phenomena at play during the formation of  different  directional eutectic melting patterns, including an unprecedented morphological transition from period-preserving to  period-doubling shapes. Based on the scaling behaviors of characteristic parameters, our analysis brings clear evidence that the large-scale shape of a eutectic melting pattern is essentially determined by a local diffusive coupling at the trijunction at large $V_m$, while it is strongly influenced by the thermal gradient when $V_m$ decreases below a critical value. Each of these phenomena  deserve further investigations. Some open questions remain to be addressed that concern, e.g., solid-state diffusion,  departure from local equilibrium  at interfaces, and 2D vs. 3D geometries, and their effect on the eutectic melting dynamics.  In view of the importance of melting in additive manufacturing, we hope that our work encourages new experimental and numerical studies.

\section*{Acknowledgments} 

We are grateful to M. Plapp and A. Karma for insightful discussions. We thank Bumedijen Raka for technical help. This work was financially supported by the collaborative, DFG and ANR  project  DYNAMELT (ANR-23-CE08-0035-01, DFG-530167777).

%
%
%
%

\clearpage


\section*{Supplemental Information}

\renewcommand{\thefigure}{S\arabic{figure}}
\setcounter{figure}{0}



\begin{figure*}[htbp] 
\includegraphics[keepaspectratio, width=12cm]{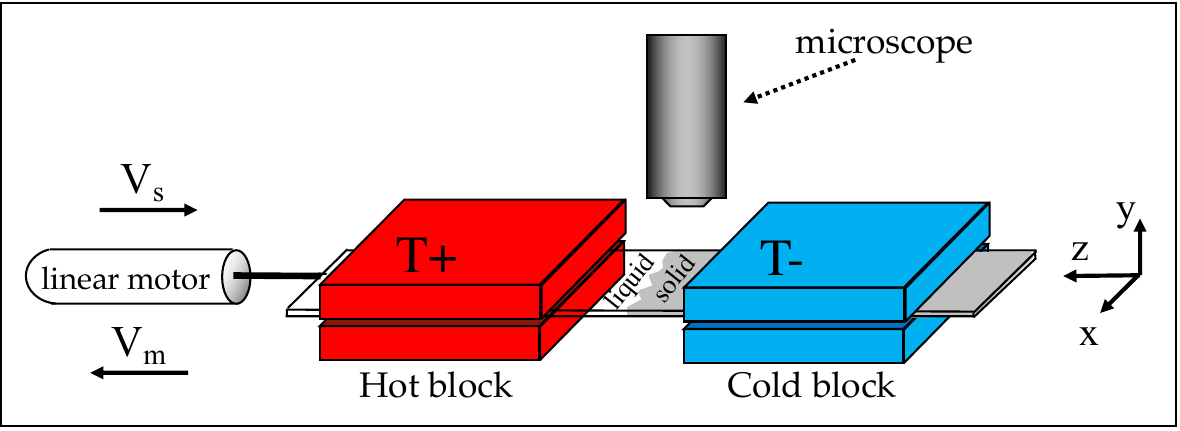}
\caption{Schematic representation of the thin-sample directional solidification method.}
  \label{TDS}
\end{figure*}

\begin{figure*}[htbp] 
  \includegraphics[keepaspectratio, width=0.6\textwidth]{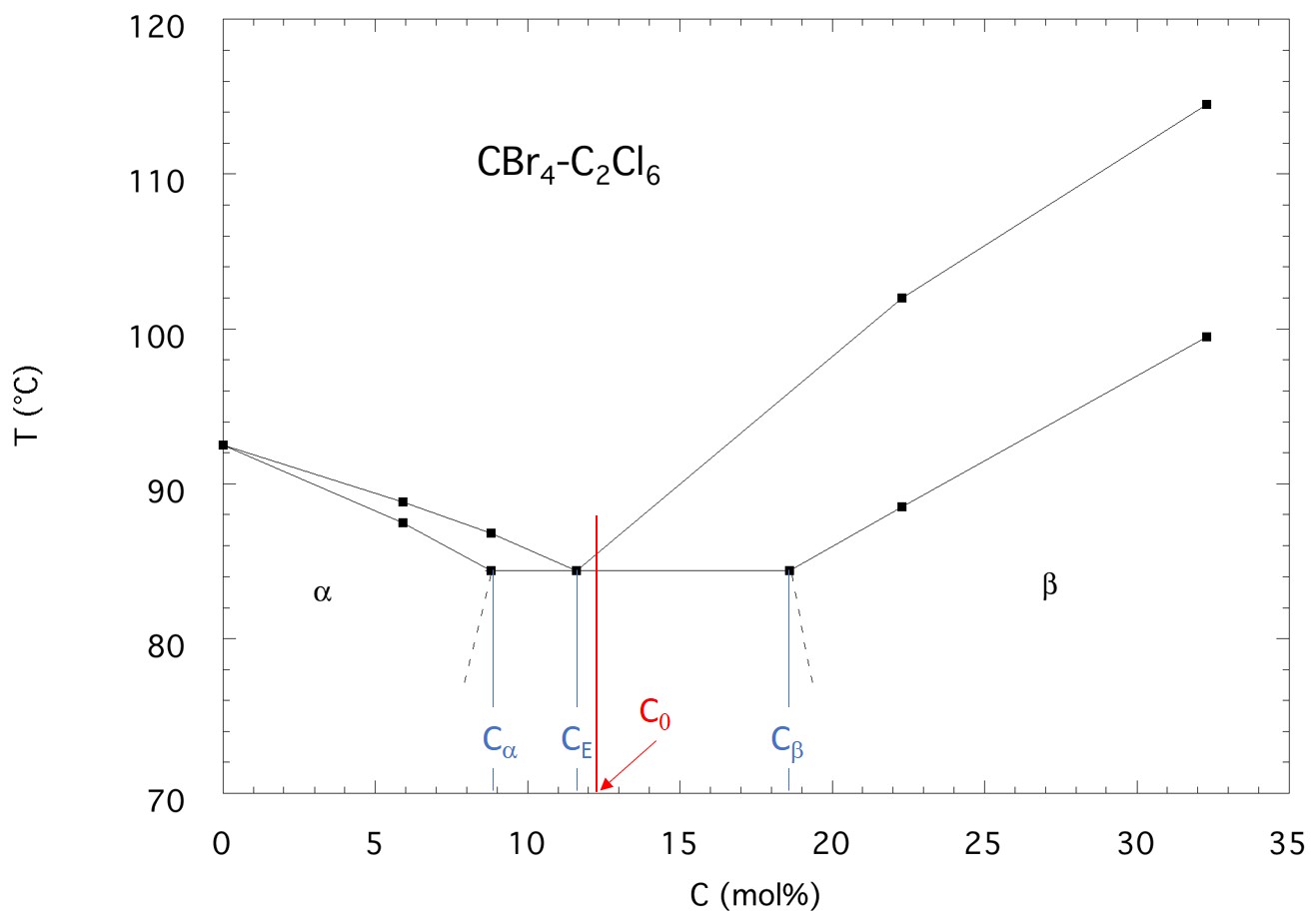}
  \caption{Phase diagram of the CBr$_4$-C$_2$Cl$_6$ eutectic.
See J. Mergy, G. Faivre, C. Guthmann, R. Mellet, Journal of
Crystal Growth 134, 353 (1993). }
  \label{phase_diag}
\end{figure*}

\begin{figure*}[htbp] 
\includegraphics[keepaspectratio, width=12cm]{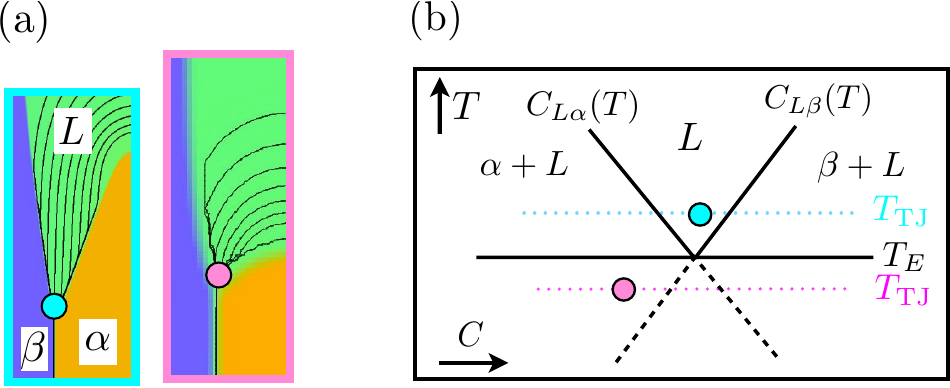}
\caption{(a) focus on the trijunction (TJ) with iso-concentration lines in the liquid for the situation $\rho \ll \lambda$ (blue) and $\rho \sim \lambda$ (pink); (b)  focus on the eutectic point in the phase diagram: the horizontal dotted lines schematically indicate the temperature $T_\text{TJ}$ of the TJ, and the circles indicate the concentration in the liquid at the TJ. }
  \label{TJ}
\end{figure*}

\end{document}